# Photonic assisted light trapping integrated in ultrathin crystalline silicon solar cells by nanoimprint lithography

Christos Trompoukis[1,2], Ounsi El Daif[1], Valérie Depauw[1], Ivan Gordon[1], Jef Poortmans[1,2]

[1] Imec, Kapeldreef 75, B-3001 Leuven, Belgium

[2] Katholieke Universiteit Leuven, Departement Elektrotechniek – ESAT, Kasteelpark Arenberg 10, B-3001 Leuven, Belgium

## ABSTRACT

We report on the fabrication of two-dimensional periodic photonic nanostructures by nanoimprint lithography and dry etching, and their integration into a 1-µm-thin mono-crystalline silicon solar cell. Thanks to the periodic nanopatterning, a better in-coupling and trapping of light is achieved, resulting in an absorption enhancement. The proposed light trapping mechanism can be explained as the superposition of a graded index effect and of the diffraction of light inside the photoactive layer. The absorption enhancement is translated into a 23% increase in short-circuit current, as compared to the benchmark cell, resulting in an increase in energy-conversion efficiency.

Most of the work on photovoltaic (PV) solar energy is currently directed towards a decrease in cost per watt of electricity produced. In order to achieve that, many technologies based on different materials are being investigated. However, crystalline silicon (c-Si) still remains the dominant technology. In fact, crystalline silicon technology presently has a market share of more than 80% and it is expected to remain the most prominent PV technology for at least the next decade. The reason lies in the fact that c-Si PV has numerous assets to enable large scale deployment. However, for present technologies which utilize 180 μm thick wafers, the material itself accounts for around 30-40% of the price of a silicon module [1]. Using less crystalline silicon material could reduce the cost. Therefore, thin-film c-Si solar cells could offer the possibility to achieve high energy conversion efficiencies by exploiting the material's high quality, at low cost.

Technologies that enable the fabrication of thin c-Si layers in the range of 1 μm [2] up to 40 μm [3, 4] directly on glass have been proposed. However, c-Si thin-films of a few micrometers suffer from a significantly reduced optical absorption in the red and near-infrared regions of the solar spectrum. In fact, the indirect band-gap of silicon yields a low absorption coefficient and the film thickness leads to a limited optical path length, which is the main disadvantage of thin-film c-Si solar cells.

In order to improve the light absorption, standard wafer-based technologies usually exploit phenomena based on geometrical optics. Random pyramid texturing, obtained through chemical wet etching of silicon, is the state-of-the-art front side texturing technique for crystalline silicon wafers [5]. Its effect is based on the refraction of light, multiple reflections and total internal reflection [5]. Combined with an anti-reflection coating (ARC) and a metallic back reflector, it yields an integrated reflectance of approximately 5-6% between wavelengths of 300-1200 nm. Despite its effectiveness, random pyramid texturing cannot be implemented in thin-film technologies because the pyramids have dimensions of a few microns and the fabrication process consumes 5 to 10 μm on each wafer side. Therefore, for thin-film c-Si solar cells, light trapping schemes with submicron dimensions are needed which could achieve efficient coupling of the incoming light to the photoactive layer.

Recent progress in nanophotonics has enabled the manipulation of light using periodic photonic structures with dimensions at the nanoscale, which is the order of magnitude of visible light wavelengths (0.1 to 1 μm) [6]. More specifically, for applications to photovoltaics it has been shown that trapping light beyond the commonly accepted limit of Lambertian light scattering is possible through the use of periodic photonic nanostructures [7-8]. The optical properties of such structures have been widely studied theoretically [9-13] and their integration has been mainly tried in a-Si, micromorph and organic based cells [14-16]. For the case of mono-crystalline silicon, periodic photonic nanostructures have been barely investigated optically but they have not been integrated in a solar cell so far [17-22].

In this paper, we report on the integration of two dimensional (2D) periodic photonic nanostructures into a 1 μm thin-film c-Si solar cell structure. The solar cells were fabricated on c-Si films obtained by a layer-transfer process and the nanostructures were patterned by nanoimprint lithography (NIL) [23] and dry-etching. After an overview of the solar cell fabrication and characterization processes, we present the topography of the nanopatterned cells and discuss about their optical and electrical performance.

As the starting substrate, mono-crystalline p-doped silicon films of 1 μm are used (photoactive layer including 20 nm of $p^+$-doped silicon used as a back surface field (BSF)).

Their thickness makes them an appropriate vehicle for investigating the optical effects from the 2D periodic photonic nanostructures. Their fabrication technique is based on the reorganization at high temperature of macroporous silicon creating a suspended void-free layer of c-Si. An aluminum (Al) layer of 1 μm is then deposited through electron-beam assisted evaporation, and this stack is then transferred and bonded to a glass substrate of 0.5 mm thickness. This "Epifree" (epitaxy-free) technique and its specific use in solar cells are described extensively in ref. [24].

The definition of the 2D periodic photonic nanostructures is done by soft thermal-NIL (Figure 1) [18]. The concept is based on the direct deformation of a thermoplastic resist by compression from a soft polydimethylsiloxane (PDMS) stamp. The PDMS soft stamp is molded using a silicon master stamp, which is patterned by deep ultraviolet (DUV) lithography and dry etching. The thermoplastic resist is a commercial nanoimprint resist from microresist technology. The imprint is performed in a home-made hydraulic press at 130°C with an imprinted area of 9 cm$^2$. Subsequently to the imprint process, a reactive ion etching (RIE) step is performed (combination of $SF_6$ and $O_2$ gases at low pressure), in order to transfer the pattern to the sample using the imprinted polymer as the mask. Finally, the resist residues are removed by dissolution in acetone.

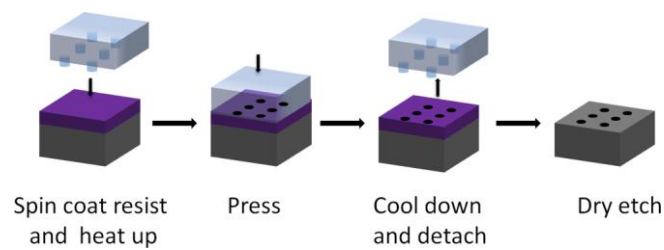

Figure 1. Schematic representation of the soft thermal-nanoimprint lithography process flow.

For the solar cell fabrication, a mesa heterojunction-emitter cell structure is used [24]. Intrinsic (10 nm) and n$^+$-doped (10 nm) hydrogenated amorphous silicon (a-Si) layers are deposited by plasma-enhanced chemical vapor deposition (PECVD) on the p-type photoactive layers in order to passivate the surface and form the emitter, respectively. In order to allow for passivation of the whole cell surface, the a-Si layers are deposited after the nanopatterning is performed. Tin-doped indium oxide (ITO) is deposited on the samples (used as anti-reflection coating (ARC) and charge collecting layer) using a sputtering tool. The emitter (n$^+$-type) contacts are electron-beam evaporated through a shadow mask directly on the ITO. Titanium/palladium/silver (Ti/Pd/Ag) is used for contacting the emitter. Silicon is finally etched down by RIE for cell separation and rear-contact opening. The final cell structure is shown in Figure 2.

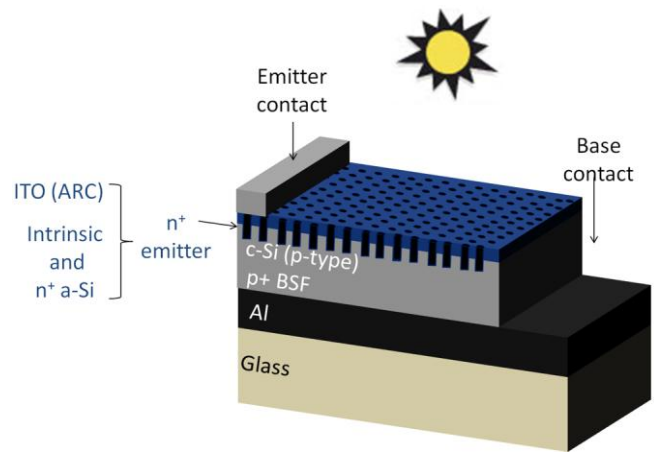

Figure 2. Schematic representation of the cell structure.

The topography was characterized with scanning electron microscopy (SEM). The optical properties of the samples were characterized by spectrally resolved reflectance (R) and transmittance (T) measurements, within a wavelength range from 300 to 1200 nm and integrated over the whole half space, using an integrating sphere. The absorption of the full stack is then given by A=1-R as there is no transmission through the metal back reflector. The illuminated current-voltage (IV) measurements were performed under AM1.5 illumination with an aperture area of 1 cm$^2$. The integrated values of absorption and reflectance are, respectively, the proportion of the solar photons absorbed and reflected over the range from 300 to 1200 nm taking into account the AM1.5 global tilt intensity distribution.

Figure 3 shows the topography of the nanopatterned 1 µm c-Si after NIL and dry etching. The period of the structures is 900 nm, the diameter is 800 nm with a depth of 550 nm. The topography of the periodic nanopattern plays an important role regarding both its optical properties and its integration in a cell fabrication process flow. On one hand, it is important to minimize the front reflection at the air/c-Si interface and trap light inside the photoactive layer. On the other hand, it is essential not to compromise the surface passivation properties of the material when texturing its surface in order to maintain high voltage values. For that reason it is critical to maintain a balance with a pattern which could offer beneficial optical behavior and, at the same time, a topography that can be efficiently passivated.

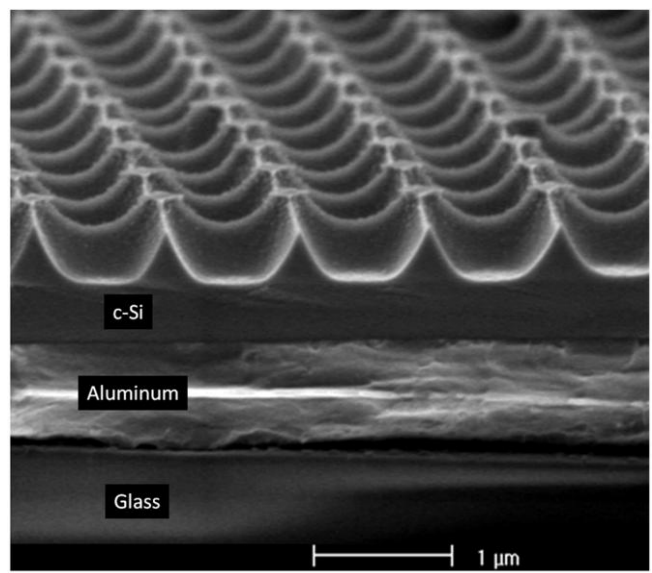

Figure 3. SEM image of the nanopatterned 1 µm c-Si sample after NIL and dry etching showing the topography of the periodic patterning and its uniformity.

It should be noted that we use a flat structure as reference because this type of cells could not be textured so far, due to the lack of alternative texturing techniques for these ultra-thin mono-crystalline silicon layers. The reflectance and absorption results of the reference and nanopatterned cells (full stack including the metal front contacts, the ARC, the a-Si emitter layer and the metal back reflector) are shown in Figure 4. The integrated reflectance of the patterned stack is 16% whereas the unpatterned stack reaches 44%. Although nearly half of the photoactive layer was removed during etching, the periodic nanopatterning results in an enhancement in absorption. The integrated absorption is increased from 53% for the unpatterned cell up to 81% for the nanopatterned cell.

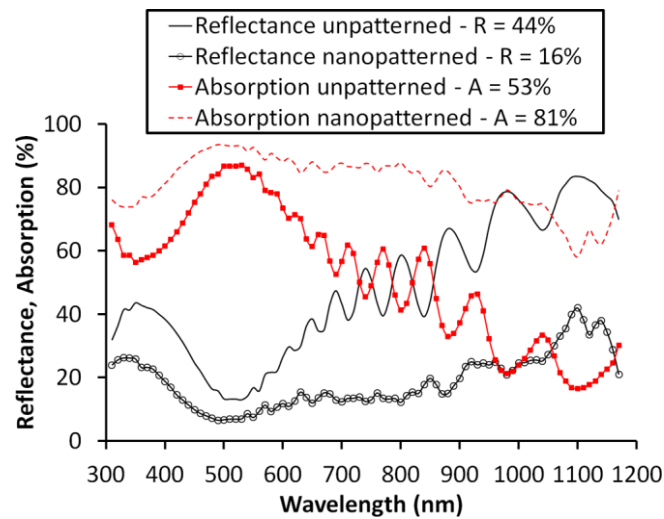

Figure 4. Lower reflectance and enhanced absorption of the nanopatterned cells as compared to the unpatterned cells (full stack including the front metal contacts, the ARC, the a-Si and the metal back reflector).

The light-trapping mechanism we propose can be understood as the superposition of two phenomena [18]. On one hand, the decrease in the front reflectance for the nanopatterned cell, which can be observed in Figure 4 at the wavelengths between 300 and 500 nm, can be attributed to a better coupling of light inside the photoactive layer with no influence of the back side. Indeed, silicon has a high absorption coefficient at this wavelength range and in-coupled light is fully absorbed before it can reach the back of the structure. For those wavelengths, diffraction takes place on the tips and walls between two holes, which have dimensions much below the wavelengths of light. On the other hand, the oscillations present for the unpatterned cell, which can be seen above 500 nm, indicate a Fabry-Perot resonance behavior. At this wavelength regime, light reaches the back of the structure and this corresponds to a low absorption case, where the silicon layer acts as a cavity containing a standing-wave. When comparing the absorption of the patterned and unpatterned structures in Figure 4 beyond 500 nm, one can observe certain elements revealing that diffraction inside the structure is taking place. Oscillations corresponding to light which is incoupled to waveguide modes [13], appear at higher wavelengths for the nanopatterned cell thanks to the periodic patterning. Moreover, these oscillations are attenuated while having a spectral dependence which is distorted. Additionally, instead of an abrupt change in the refractive index (RI) when passing from air to c-Si, light goes through a progressive change due to the non-vertical sidewalls of the pattern, starting from a tip which is much smaller than the wavelength, allowing a progressive increase of the silicon fraction in the surface seen by incident light (graded index effect). Overall, the RI mismatch from the air/c-Si interface is improved by the combination of diffraction of light inside the photoactive layer and a graded index effect, which lead to an absorption enhancement.

Since the optical results refer to the full cell stack, the measured spectra include parasitic absorption and reflectance originating from materials other than the photoactive layer. The ITO and a-Si layers absorb part of the incident light for short wavelengths. Additionally, there is a shadowing effect due to the presence of the front metal contacts. Finally, the Al back reflector also contributes to parasitic absorption [17]. These effects can be discriminated by the electrical measurements, where only photons absorbed in the active layer contribute to the photocurrent.

The electrical results show the translation of the optical improvement into a current enhancement, while not significantly affecting the remaining electrical parameters of the reference cell. As can be seen on the illuminated current-voltage (IV) curves (Figure 5a), the absorption enhancement which is observed from the optical characterization is confirmed and translated into an increase in short-circuit current ($J_{sc}$). The spectral response of the cells is shown in Figure 5b and the electrical results are summarized in Table I. The values in that table refer to the average values from three cells, while the best obtained cell is mentioned in brackets.

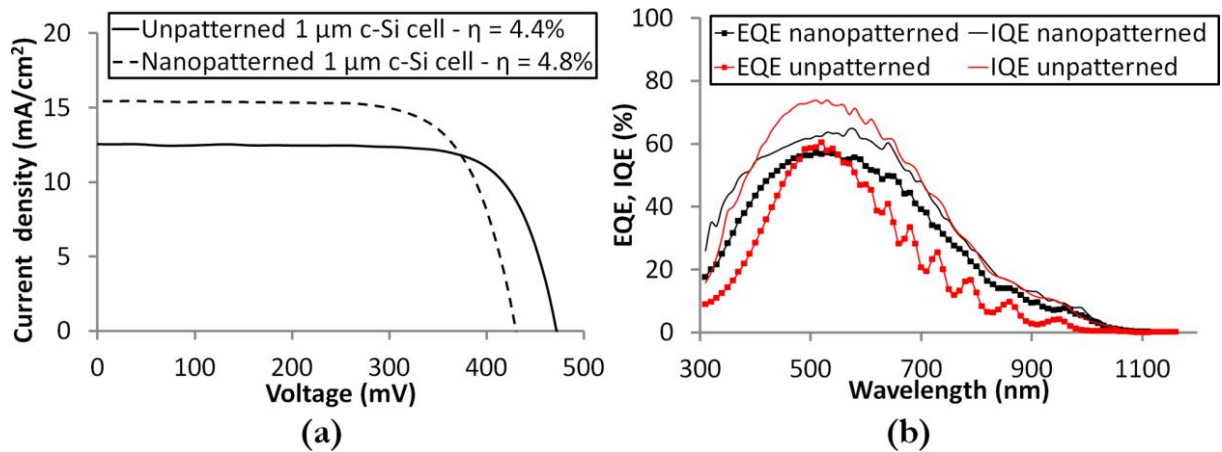

Figure 5. a) Illuminated current-voltage characteristic and b) spectral response of the best nanoimprinted 1 μm c-Si cell highlighting the enhancement in current compared to the best unpatterned cell.

TABLE I. Performance parameters for the nanopatterned and the unpatterned cells (values for the best cells in the parenthesis)

| Texturing | $J_{sc}$ [mA/cm$^2$] | $V_{oc}$ [mV] | FF [%] | η [%] |
|---|---|---|---|---|
| No | 12.6 (12.5) | 471 (471) | 72 (75) | 4.3 (4.4) |
| 2D periodic nanopatterning | 15.3 (15.5) | 435 (434) | 72 (72) | 4.8 (4.8) |

The $J_{sc}$ is increased by 23% thanks to the presence of the periodic nanopattern, from 12.6 to 15.3 mA/cm$^2$ (Table I). As shown in Figure 5b, the external quantum efficiency (EQE) of the nanopatterned cell is higher than that of the unpatterned cell on most part of the wavelength range. More precisely, for short (below 450 nm) and long (above 550 nm) wavelengths the EQE of the nanopatterned cells is higher, highlighting the reduced front reflection and the enhanced light trapping compared to the unpatterned cells, respectively. The observed drop in voltage may be related to an increase in the minority-carrier recombination velocity, due to the increased surface area and to damage from the ion bombardment during the RIE. In particular, it has been shown [25] that the damage due to the RIE can reach a depth of 1 µm which, in our case, is more than the remaining photoactive layer. This is also confirmed by Figure 5b showing that the internal quantum efficiency (IQE) of the nanopatterned cells is lower (from 400 to 650 nm), which suggests that both surface and bulk lifetimes were affected. However, despite this damage, the open-circuit voltage ($V_{oc}$) value decreased by only 8% (Table I). This resulted in an overall improvement of the energy-conversion efficiency of the nanopatterned cell as compared to the unpatterned cell, increasing from a value of 4.3 to 4.8%. Therefore, this shows that the impact of the nanopattern on the passivation of the cells was sufficiently limited while it could significantly enhance the photon collection.

In the future, the efficiency could further increase both from improvements in the periodic nanopattern and in the cell design. More precisely, optical simulations indicate that an optimized set of parameters for the nanopattern could further enhance the effect leading to higher $J_{sc}$ values [26]. Additionally, a cell structure including an improved back-surface field, an optical spacer adjacent to the back reflector as well as improved passivation and contacting schemes would further increase the $V_{oc}$ and the fill factor values.

In conclusion, we have demonstrated the integration of a nanopatterning step in the process flow of a 1-µm-thin c-Si solar cell. The nanopatterning was performed by nanoimprint lithography and reactive ion etching, and resulted in a minimal thickness loss of less than 600 nm and an overall absorption enhancement. The resulting absorbance indicates that the light trapping mechanism is based on the superposition of a graded index effect and the diffraction of light inside the photoactive layer. The device performance showed that the nanopatterning was beneficially integrated, and that it improved the optical properties of the solar cell without heavily affecting the electrical and material properties. Although almost half of the initial photoactive layer was removed during the nanopatterning, this resulted in an enhanced absorption and an increase of 23% of the short circuit current. Therefore, we demonstrate that light management features at the wavelength scale can be integrated in a thin-film cell structure enhancing its efficiency. The optimization of the nanopattern parameters and its integration in a more elaborate solar cell fabrication flow should allow a further enhancement of the cell efficiency. This could offer a solution for a cost effective photovoltaic technology thanks to the use of photonic assisted thin-film solar cells.